\documentclass[amsmath,nofootinbib,amsfonts,amssymb,aps,prd,notitlepage]{revtex4-1}

\usepackage{hyperref}
\usepackage[english]{babel}
\usepackage{graphicx}
\usepackage{bm}
\usepackage{slashed}
\usepackage{color}
\usepackage{multirow}
\usepackage{array}
\usepackage{hhline}

\begin{document}

\title{Discovering sterile Neutrinos ligther than $M_W$ at the LHC}
\author{Claudio O.~Dib\footnote{claudio.dib@usm.cl} }
\affiliation{ CCTVal and Department of Physics,
Universidad T\'ecnica
Federico Santa Mar{\'{\i}}a, Valpara{\'{\i}}so, Chile}
\author{C. S. Kim\footnote{cskim@yonsei.ac.kr} }
\affiliation{ Department of Physics and IPAP, Yonsei University, Seoul 120-749, Korea}

\date{\today}

\begin{abstract}
\noindent We study the purely leptonic $W$ decays $W^+ \to e^+ \mu^- e^+  \nu_e$ and $W^+ \to e^+ e^+ \mu^- \bar \nu_\mu$ (or their charge conjugates) produced at the LHC, induced by sterile neutrinos with mass below $M_W$ in the intermediate state. While the first mode is induced by both Dirac or Majorana neutrinos, the second mode is induced only by Majorana neutrinos,
as it violates lepton number.  We find that, even when the final (anti-)neutrino goes undetected, one could distinguish between these two processes, thus distinguishing the Dirac or Majorana character of the sterile neutrinos, by studying the muon spectrum in the decays.
\end{abstract}


\maketitle

\section{Introduction}

One of the outstanding issues in neutrino physics today is to
clarify the  Dirac or Majorana character of neutrino masses.
The discovery of neutrino oscillations indicates that neutrinos
are massive particles, albeit with masses much smaller than
those of the charged fermions \cite{Agashe:2014kda}.
This evidence may provide an important clue to the existence of a more
fundamental physics underlying the Standard Model (SM). Neutrinos are naturally massless in the
SM, with the consequent conservation of all lepton flavors -- electron, muon and tau -- separately. On the other hand, massive neutrinos
could naturally induce violation of $e$, $\mu$ and $\tau$ flavors (due to flavor mixing analogous to the quark sector) and also induce  violation of total lepton number, due to the appearance of Majorana masses in the neutrino sector.  Indeed, in the simplest scenario of neutrino masses, these are due to Yukawa interactions -- the mechanism that gives mass to all other fermions. However, this scenario requires the inclusion of right-handed neutrino fields, which in turn allows Majorana mass terms for these right-handed fields
~\cite{Minkowski:1977sc, GellMann:1980vs, Yanagida:1979as, Mohapatra:1979ia, Glashow:1979nm, Schechter:1980gr}.
The neutrino sector then becomes richer and, through mixing, some neutrino masses become small as others become large, an effect called \emph{seesaw}. To date, there are many versions of seesaw models, all of them with the common feature of having a spectrum separated into light and heavier neutrinos, with their corresponding mixing to the lepton currents of the electroweak interactions (see e.g. \cite{Ma:2009dk, FileviezPerez:2009ud} and references therein).

In seesaw models based on the SM gauge group, the heavy neutrinos are coupled to the standard sector through a small mixing with the standard leptons in the electroweak currents. Here we will focus on these cases.  Scenarios based on other gauge groups such as left-right symmetric gauge theories, where  the heavy neutrinos connect to the standard sector primarily through right-handed currents, will not be considered here. This is important to keep in mind, because in the first case the couplings with extra neutrinos are suppressed by small mixings, while in left-right symmetric or other gauge groups the suppression may instead come from large masses of the extra gauge bosons, such as $W_R$~\cite{Keung:1983uu}.
Each specific scenario proposes heavy neutrinos with masses within a given scale, but in general this scale can be in a very broad range, from a few
eV all the way to grand unification scales of order $10^{15}$ GeV.
In turn, different experiments put bounds on neutrino masses and mixings, each one in a different and limited mass range within this broad spectrum of possibilities~\cite{Atre:2009rg}.

The most sensitive processes considered to search for the Majorana character of neutrinos are neutrinoless nuclear double beta decays ($0\nu\beta\beta$) \cite{Maneschg:2015dja, Simkovic:2013kna, Bilenky:2012qi, Faessler:2009zz}.  A positive signal will clearly indicate that neutrinos are of Majorana type, even though the extraction of neutrino parameters will be difficult, especially  due to large theoretical uncertainties in the nuclear matrix element.
Heavy neutrinos, either Dirac or Majorana, may also give observable effects in rare meson decays~\cite{Cvetic:2010rw, Cvetic:2012hd, Lees:2013gdj, Liventsev:2013zz, Bonivento:2013jag, Dib:2014iga} or in high energy collisions~\cite{Deppisch:2015qwa, Abada:2013bpa, Das:2012ze, Das:2014jxa}, where lepton flavor and/or lepton number are violated~\cite{Keung:1983uu, BarShalom:2006bv,  delAguila:2008cj}. In particular at the LHC one looks for events with two same-sign leptons together with  two jets and no missing energy 
\cite{ Aad:2015xaa, Khachatryan:2015gha}, or  three charged leptons and missing energy, with appropriate cuts to eliminate backgrounds. For neutrino masses below $M_W$, the heavy neutrino could live long enough as to cause a visible
displacement between the vertex where the neutrino is produced and where it decays \cite{Helo:2013esa, Izaguirre:2015pga, Gago:2015vma}, a feature that can be used in order to eliminate backgrounds. For semileptonic modes $\ell^\pm\ell^\pm j j $ this displacement is considered  in Ref.~\cite{Helo:2013esa}. 
However, in these semileptonic modes, to avoid severe hadronic background and soft pile up processes at the LHC, the cuts that are imposed to keep only hard and well isolated jets make it difficult to observe events with intermediate on-shell neutrino masses  below  $M_W$. 
A more appropriate process for  $m_N < M_W$ could be the purely leptonic mode  $W^+ \to e^+ N(\to e^+ \mu^- \bar\nu_\mu)$ or its conjugate \cite{Izaguirre:2015pga},  since charged leptons are easier to identify than jets at lower transverse momenta. One should also notice that these leptonic on-shell
$W$ decays can be approximately fully reconstructed event-by-event, albeit with two-fold ambiguity \cite{Cortes:1985yi, Halzen:2013bqa} at the LHC, if only a single --almost massless-- neutrino escapes detection in the $W$ decay.

Here we study these leptonic $W$ decays 
in order to address the discovery of heavy sterile neutrinos 
at the LHC
with masses below $M_W$ and distinguish their Majorana vs.\ Dirac character.
Specifically, we want to consider the LNV decays $W^+ \to e^+ e^+ \mu^- \bar\nu_\mu$ \cite{Izaguirre:2015pga}, and
the lepton number conserving (LNC) decays $W^+\to e^+ e^+ \mu^- \nu_e$.  One can just as well consider their charge conjugate processes but, from now on and just for the sake of notation, we will refer to the indicated process, keeping in mind that their conjugates can be used as well.
If $m_N < M_W$, these processes are dominated by a resonant heavy neutrino in the intermediate state, namely
$W \to e N(\to e \mu \nu)$.
These leptonic decays violate lepton flavor, and in the case of $W^+ \to e^+ e^+ \mu^- \bar\nu_\mu$  it violates lepton number as well, with $\Delta L=2$ just as the semileptonic decay $W^\pm \to \ell^\pm N(\to  \ell^\pm~ j j)$ mentioned above.
However, in the purely leptonic case
the final neutrino flavor goes undetected, and therefore the LNC (albeit lepton flavor violating) process
$W^+ \to e^+ \mu^- e^+ \nu_e $ cannot be separated from the $\Delta L=2$ mode.
Moreover, while both processes are mediated by a heavy Majorana neutrino, only the second process occurs if it is mediated by a Dirac neutrino.
It is then important to distinguish these two modes if we want to test the Majorana $vs.$ Dirac character of the neutrino.

Here we will explore how to distinguish them and thus test the existence of the $\Delta L= 2$ process.
We must also note that, by specifically choosing $e^\pm e^\pm \mu^\mp$ (or $\mu^\pm \mu^\pm e^\mp$)
 instead of $e^\pm \mu^\pm e^\mp$ in the final state,
one avoids serious backgrounds,
from $e.g.$ radiative decays $W^+ \to \mu^+ \nu_\mu + \gamma^*(\to e^+ e^-)$ \cite{Cvetic:2012hd}. 
In Section II we present the details of the scenario with formulae for the rates and spectra of the decays under consideration, and in Section III we give a summary and conclusions.

\section{Rates and Spectra of $W^\pm \to e^\pm e^\pm \mu^{\mp} \nu$ at the LHC}


 We want to study lepton flavor violating leptonic decays of an on-shell $W$, mediated by a heavy neutrino with mass below $M_W$. Moreover, if the intermediate neutrino is of Majorana type, the process will also exhibit lepton number violation.
 The LNV process, $W^+\to e^+ e^+ \mu^- \bar\nu_\mu$, is depicted in
 Fig.~\ref{fig:muee}. Since the final (anti-)neutrino goes undetected experimentally, one should also consider the process $W^\pm\to e^+ e^+ \mu^- \nu_e$ that has the same charged leptons in the final state and violates lepton flavor, but conserves lepton number. This process is shown in Fig. 2. It is important to notice that an intermediate neutrino of Majorana type will produce both processes, while an intermediate neutrino of Dirac type will only produce the second process.
Given that the final neutrino is not detected, an important question that arises is whether one can actually distinguish these two processes, otherwise they can not be used to determine the Dirac or Majorana character of the intermediate neutrino.
We will show that these two decays can actually be separated at least in the spectrum, if not in the rate.

\subsection{The Lepton Number Violating 
Process $W^+ \to \ell^+ \ell^+ \ell^{\prime -} \bar\nu_{\ell^{\prime}}$}

\begin{figure}[h]
\includegraphics[scale=0.5]{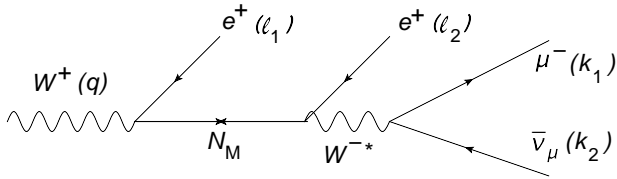}
\caption{The diagram for the lepton number violating (LNV) decay $W^+ \to e^+e^+\mu^-\bar\nu_\mu$, which is mediated by a Majorana  neutrino.}
\label{fig:muee}
\end{figure}

Let us first consider the LNV decay  $W^+ \to e^+ e^+ \mu^-\bar\nu_\mu$ (or equivalently $W^+ \to \mu^+ \mu^+ e^- \bar\nu_e$ --or their charge conjugated decays), shown in Fig.~\ref{fig:muee}. There is only one diagram at tree level if there is no flavor-changing neutral current in the $Z$ coupling ($i.e.$ there is no vertex $Z\to \mu^- e^+$), an assumption we will use in this work.

The squared matrix element for the process in Fig.\ref{fig:muee}, averaged over initial polarizations, is:
\begin{equation}
 |{\cal\overline  M}|^2 = 256 \frac{\sqrt{2}}{3}   G_F^3 M_W^2   |U_{N e}|^4  \  \frac{1}{(k_N^2-m_N^2)^2+m_N^2\Gamma_N^2}  \  m_N^2 (k_2 \cdot \ell_2)  \Big\{ ( k_1 \cdot \ell_1) + \frac{2}{M_W^2}  (q\cdot k_1 )  (q\cdot \ell_1)
   \Big\},
\label{mesq}
\end{equation}
where $k_N = q - \ell_1$ is the virtual momentum of the intermediate neutrino $N$, which will be predominantly on its mass shell when $m_N$ is in our range of interest, due to the sharply peaked neutrino propagator.
Indeed, since the intermediate neutrino is weakly interacting, we can clearly use the narrow width approximation for its propagator, which is to say that the neutrino is produced as an on-shell particle which  subsequently decays. For this purpose one can express the 4-particle final phase space as:
%
%
\[
 \int d_{ps4}(q\to\ell_1,\ell_2, k_1, k_2) =
 \int \frac{d k_N^2}{2\pi}
 \int d_{ps2} (q\to\ell_1, k_N)
 \int d_{ps3} (k_N\to k_1, k_2, \ell_2 )
%
\]
Provided $m_N$ is such that $m_\mu + m_e < m_N < M_W - m_e$, which is the range of interest here, the narrow neutrino width allows us to approximate:
\[
\int \frac{d k^2}{2\pi} \frac{1}{(k^2-m_N^2)^2 + m_N^2 \Gamma_N^2} \to \frac{1}{2 m_N \Gamma_N},
\]
thus fixing $k_N$ to be on its mass shell.
%
%
%
%
Doing the phase space integrals we can get the $\mu^-$ spectrum in the $N$ rest frame:
\begin{eqnarray}
\Gamma (W^+\to e^+ e^+\mu^-\bar\nu_\mu)&=&
\frac{G_F^3 M_W^3}{12 \sqrt{2} \ \pi^4}  \frac{ |U_{N e}|^4 m_N}{\Gamma_N}
\left( 1-\frac{m_N^2}{M_W^2} \right)^2    \left(1 + \frac{m_N^2}{2 M_W^2} \right)
\nonumber\\
&& \hspace{1cm} \times
\int_0^{m_N/2} d E_{\mu}
  \left( m_N E_{\mu}^2 - 2 E_{\mu }^3\right)
  .
  \label{spectrum1}
\end{eqnarray}
Integrating this expression we can obtain the branching ratio:
%
%
%
\begin{eqnarray}
Br(W^+\to e^+ e^+\mu^-\bar\nu_\mu)
&=&
\frac{1}{12\times 96 \pi} \left(\frac{G_F}{\sqrt{2}} \frac{M_W^3}{ \Gamma_W}\right)
 \left(
 |U_{N e}|^4 \frac{G_F^2   \, m_N^5}{\pi^3 \Gamma_N}
 \right)
 \
\left(1 - \frac{m_N^2}{M_W^2}\right)^2
 {\left( 1 + \frac{m_N^2}{2 M_W^2}\right) }
  \nonumber\\
 \label{bratio}
%
 &\approx&
 4.8 \times 10^{-3} \  \frac{|U_{Ne}|^4}{\sum_{\ell=e,\mu ,\tau} |U_{N\ell}|^2}
 \left(1 - \frac{m_N^2}{M_W^2}\right)^2
 {\left( 1 + \frac{m_N^2}{2 M_W^2}\right) }  .
\end{eqnarray}

\subsection{The Lepton Number Conserving 
Process $W^+ \to \ell^+ \ell^{\prime -} \ell^+ \nu_\ell$}

\begin{figure}[h]
\includegraphics[scale=0.5]{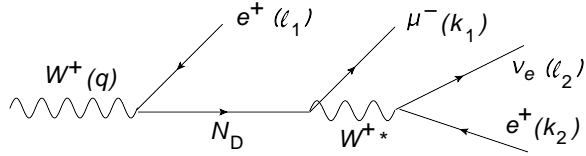}
\caption{The diagram for the lepton number conserving (LNC) decay $W^+ \to e^+ \mu^- e^+ \nu_e$, which is mediated by either a Dirac or Majorana neutrino.}
\label{fig:muee_bk}
\end{figure}

%
Now, concerning the LNC decay $W^+ \to e^+ \mu^- e^+ \nu_e$ (or equivalently $W^+ \to \mu^+  e^- \mu^+ \nu_\mu$ --or their charge conjugated decays), the diagram for its amplitude is shown in Fig.~\ref{fig:muee_bk}.

%
The corresponding squared matrix element, averaged over initial polarizations, now is:
\begin{eqnarray}
 |{\cal\overline  M}|^2 &=& 256 \frac{\sqrt{2}}{3}   G_F^3 M_W^2   |U_{N e} U_{N \mu}|^2 \times  \frac{1}{(k_N^2-m_N^2)^2+m_N^2\Gamma_N^2} \\
&\times&
(k_1 \cdot \ell_2)  \Bigg\{ 2 (k\cdot k_2) \Big[ ( k \cdot \ell_1) + \frac{2}{M_W^2}  (q\cdot k )  (q\cdot \ell_1) \Big] - m_N^2 \Big[   (k_2 \cdot \ell_1) + \frac{2}{M_W^2}  (q\cdot k_2 )  (q\cdot \ell_1) \Big]
   \Bigg\}\label{mesq2},
   \nonumber
\end{eqnarray}
where the momenta are named as before.
%
Again, one finds
%
%
%
the spectrum over the $\mu^-$ energy, $E_{\mu}$, in the intermediate neutrino $N$ rest frame:
\begin{eqnarray}
\Gamma (W^+\to e^+ e^+\mu^-\nu_e)
&=&
\frac{G_F^3 M_W^3}{12 \sqrt{2} \, \pi^4}
|U_{N e} U_{N \mu}|^2
\frac{m_N}{\Gamma_N}
\left(1-\frac{m_N^2}{M_W^2} \right)^2 \left(1  + \frac{m_N^2}{2 M_W^2}\right)
\nonumber \\ 
&& \hspace{1cm} \times
\int_0^{m_N/2} d E_{\mu}
 \left( \frac{m_N}{2}  E_{\mu}^2 - \frac{2}{3}  E_{\mu}^3 \right)
,
\label{spectrum2}
\end{eqnarray}
and from here we obtain the branching ratio for this LNC decay:
%
%
\begin{eqnarray}
Br(W^+\to e^+ e^+\mu^-\nu_e)
&=&
\frac{1}{12\times 96 \pi} \left(\frac{G_F}{\sqrt{2}} \frac{M_W^3}{ \Gamma_W}\right)
 \left(
 |U_{N e} U_{N \mu}|^2 \frac{G_F^2   \, m_N^5}{\pi^3 \Gamma_N}
 \right)
 \
\left(1 - \frac{m_N^2}{M_W^2}\right)^2
 {\left( 1 + \frac{m_N^2}{2 M_W^2}\right) }
 \nonumber\\
 \label{bratio2}
%
 &\approx&
 4.8 \times 10^{-3} \  \frac{|U_{Ne}|^2 |U_{N\mu}|^2}{\sum_{\ell=e,\mu ,\tau} |U_{N\ell}|^2}
 \left(1 - \frac{m_N^2}{M_W^2}\right)^2
 {\left( 1 + \frac{m_N^2}{2 M_W^2}\right) }  .
\end{eqnarray}
This expression is almost the same as the branching ratio of the LNV decay, Eq.~\ref{bratio}, except for the lepton mixing factors, $|U_{N\ell}|$. Clearly one cannot easily distinguish between these two processes using the rate, unless one knew the heavy-to-light mixing elements beforehand and with great precision. The  only sensible way to differentiate between the LNV and LNC decays is not by their rate but by the shape of their muon spectrum, the detail of which will be shown in below.

\subsection{Numerical Results and Discussions}

The observation of any of these two decays will be indication of the existence of an intermediate sterile neutrino  with mass below $M_W$, but if in addition we want to determine the Dirac $vs.$ Majorana character of this neutrino, we must distinguish between the LNV and the LNC decays.
This distinction is necessary because
%
the LNC decay can be induced by an intermediate neutrino of either Dirac or Majorana type, while the  LNV decay can be induced only by a Majorana neutrino.
In other words, if the intermediate neutrino is of Dirac type only the LNC decay occurs, not the LNV decay, while if the intermediate neutrino is Majorana, both decays will occur, with a relative rate that depends on a ratio of lepton mixing elements.



Let us now examine the energy spectrum of the muon ($i.e.$ the opposite-charge lepton) for the LNV and LNC decays in question. To keep the expressions simple, we use the muon energy distributions in the rest frame of the neutrino $N$.  One can also boost the spectra to the $W^\pm$ frame or to the lab frame, once the $W$ momentum is reconstructed for each event, an issue that we address further below.

\begin{figure}[h]
\includegraphics[scale=0.6]{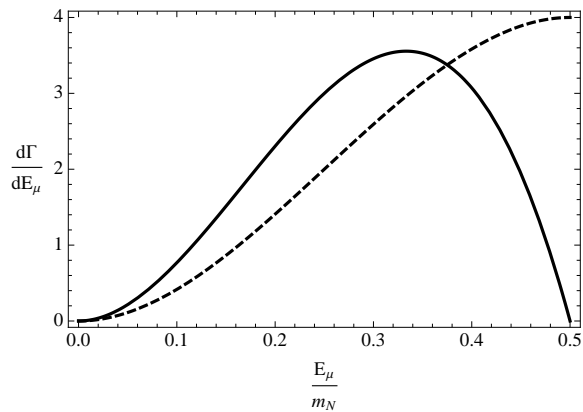}
\caption{Normalized muon energy spectrum, $(1/\Gamma)d\Gamma/dE_\mu$,  given in the $N$ rest frame, for the LNV process $W^+\to e^+ e^+ \mu^-\bar\nu_\mu$ [c.f. Eq.~(\ref{spectrum1})] (solid line),  and for the LNC process $W^+\to e^+ e^+ \mu^-\nu_e$ [c.f. Eq.~(\ref{spectrum2})] (dashed line).  In each curve the normalizing factor $\Gamma$ is the rate itself, so that the integral over either normalized spectrum is unity.}
\label{fig:spectra}
\end{figure}

The spectral shapes for the LNV and LNC decays (Eqs. \ref{spectrum1} and \ref{spectrum2}, respectively) are shown in Fig. \ref{fig:spectra}.
As seen from the figure, the two shapes are clearly different, especially at the endpoint. This is a preliminary indication that it may be possible to distinguish the two decays, but we should take into account the caveat
that,  if $N$ is a Majorana neutrino, both processes will exist, with relative rates that depend on  the heavy-to-light lepton mixing: while the LNV rate is proportional to $|U_{N e}|^4$, the LNC rate is proportional to $|U_{N e} U_{N \mu}|^2$; so far we do not know these values but in general they should be different. Consequently, if $N$ is a Dirac neutrino an experiment should find the spectrum of the LNC process; however, if $N$ is Majorana, the spectrum will not be that of the LNV process but a combination of the LNV and LNC processes.

\begin{figure}[h]
\includegraphics[scale=0.6]{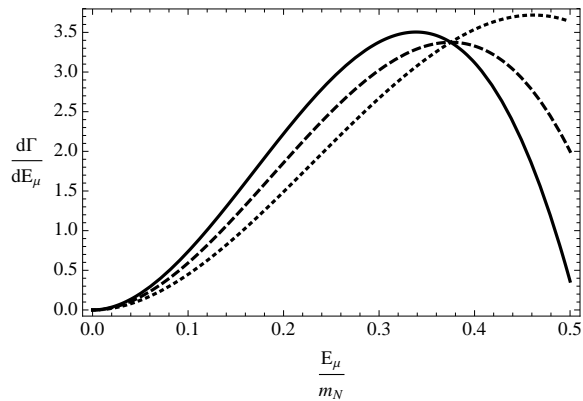}
\caption{Normalized muon energy spectrum for the LNV + LNC decays $d\Gamma/dE_\mu (W^+\to e^+ e^+ \mu^-\bar\nu_\mu) +d\Gamma/dE_\mu (W^+\to e^+ e^+ \mu^-\nu_e)$, normalised by the sum of the two rates. The rates are  proportional to $|U_{Ne}|^2$ and $|U_{N\mu}|^2$, respectively. The curves correspond to $|U_{Ne}|^2/|U_{N\mu}|^2$ $= 10, 1,$ and $ 1/10$ (solid, dashed and dotted lines, respectively). Similar shapes correspond to the spectrum of $d\Gamma/dE_\mu (W^+\to \mu^+ \mu^+ e^-\bar\nu_e) +d\Gamma/dE_\mu (W^+\to \mu^+ \mu^+ e^-\nu_\mu)$ if one exchanges the roles $|U_{Ne}|^2 \leftrightarrow |U_{N\mu}|^2$. }
\label{fig:spectra3}
\end{figure}

Let  us then consider the possible muon spectrum that an experiment would observe in the latter case, $i.e.$ $N$ is a Majorana neutrino:  both LNV and LNC processes will occur,
and the observed spectrum will be the sum of the  LNV and LNC spectra, with a proportion that will depend on the ratio $|U_{Ne}|^2/|U_{N\mu}|^2$. After normalizing, the general expression for the observed muon spectrum in the case of a  Majorana $N$ will be:
\[
\left(\frac{1}{\Gamma_{LNV} + \Gamma_{LNC}}\right)  \frac{d\Gamma}{d \varepsilon_\mu} =
\frac{1}{|U_{N e}|^2 +|U_{N\mu}|^2} \left\{ |U_{N e}|^2 \left( \varepsilon_\mu^2 - 2 \varepsilon_\mu^3 \right) + |U_{N\mu}|^2 \left( \frac{1}{2} \varepsilon_\mu^2 -\frac{2}{3}\varepsilon_\mu^3\right ) \right\} ,
\]
where $\varepsilon_\mu = E_\mu/ m_N$ is  the normalised muon energy in the $N$ rest frame. This spectrum
 is drawn in Fig.~\ref{fig:spectra3} for different values of $|U_{N\mu}|^2/|U_{N e}|^2$. The possibility to distinguish a Majorana $vs.$ Dirac neutrino clearly depends on this value.
The solid line shows the spectrum when $|U_{N e}|^2 / |U_{N \mu}|^2 = 10$, which means that the LNV mode is 10 times more intense than the LNC mode. In such case, the spectrum shows a clear drop as it reaches the endpoint, differing considerably from the spectrum corresponding to a Dirac $N$ ($i.e.$ the purely LNC spectrum) where it is maximal near the endpoint. As an intermediate case, the dashed line corresponds to $|U_{N\mu}|^2 =  |U_{N e}|^2$ (both modes with equal strength), showing that the end of the spectrum stills develops a drop, distinguishing it from the Dirac scenario.  Finally the the dotted line corresponds to $|U_{N e}|^2 / |U_{N \mu}|^2 = 1/10$, $i.e.$ when the LNC process is 10 times more intense than the LNV process, showing that the drop at the endpoint is disappearing, thus losing the capacity to distinguish between a Majorana $vs.$ a Dirac $N$. However, if this happens to be the case, what one should do is to exchange the roles of muons and electrons and study the analogous processes $W^+ \to \mu^+ \mu^+ e^- \bar\nu_e$ and $W^+\to \mu^+ \mu^+ e^- \nu_\mu$, respectively. In such case, the observed spectrum due to a Majorana neutrino is the reverse of Fig.~\ref{fig:spectra3}: the solid line corresponds to the \emph{smaller} value
$|U_{N e}|^2 / |U_{N \mu}|^2 = 0.1$ and the dotted line to the \emph{larger} value 10. The mode $W^+\to \mu^+\mu^+ e^-\nu$ thus discriminates Majorana from Dirac in a  complementary way with respect to the originally described mode $W^+\to e^+ e^+ \mu^-\nu$.

A crucial point in our analysis is to be able to reconstruct the momentum of the decaying $W$ boson, event by event. This reconstruction can actually be done to a good approximation in spite of the missing neutrino in the final state, due to the relatively small transverse momentum of the
$W$ boson \cite{Halzen:2013bqa}. The production of a on-shell $W$ at the LHC is mainly by $pp \to W^\pm X$, so that the transverse momentum of the $W$ should be of the order of the Fermi momentum of the partons inside the proton. For the leptonic decay $W\to e e \mu \nu$,
let us call $q$ the 4-momentum of the $W$, and $p$ the total 4-momentum of the three charged leptons, with invariant
mass $ p^2 \equiv \mu_{\ell}^2 $, total energy $E_{\ell}$ and total longitudinal momentum $p_L$. If we neglect the transverse momentum of the $W$, simple kinematics leads to:
%
%
\[
M_W^2  + \mu_\ell^2 - 2 E_W  E_\ell + 2 q_{L} p_{L}  =0 , \quad \textrm{where} \ \  E_W = \sqrt{M_W^2 + q_L^2}.
\]
%
This is a quadratic equation for $q_L$ (the $W$ longitudinal momentum) with solution given by:

\[
q_L = \frac{(M_W^2+\mu_\ell^2)p_L}{2 (E_\ell^2 - p_L^2)}\pm \sqrt{ \left(\frac{(M_W^2+\mu_\ell^2)p_L}{2 (E_\ell^2 - p_L^2)}\right)^2 + \frac{(M_W^2+\mu_\ell^2)^2}{4 (E_\ell^2 - p_L^2)} - \frac{M_W^2 E_\ell^2}{ (E_\ell^2 - p_L^2)}            }
\]
Since the transverse momentum of the $W$ is neglected, $q_L$ gives an approximate reconstruction of the $W$ momentum. Notice that this procedure differs from the statistical determination of the $W$ mass from leptonic decays $W\to e\nu$ at hadron colliders \cite{Abbott:1997ww}, because here $M_W$ is a known input parameter.

%

\begin{figure}[h]
\includegraphics[scale=0.8]{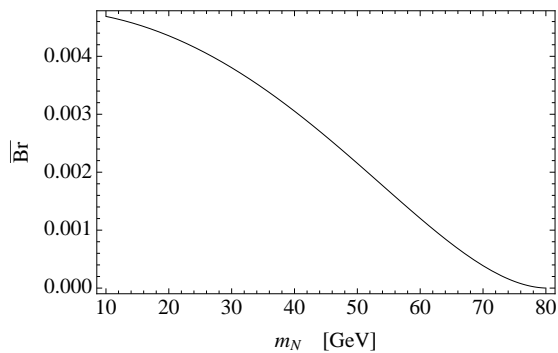}
\caption{The reduced branching ratio $\overline{Br}$ as a function of the intermediate neutrino mass $m_N$ [as in Eqs.~(\ref{bratio}) and  (\ref{bratio2})], where the lepton mixing factors have been removed.  The actual  branching ratios are obtained  as $Br = \overline{Br}\times  |U_{Ne}|^4/ (\sum_{\ell} |U_{N\ell}|^2)$ and $Br = \overline{Br}\times  |U_{Ne}U_{N\mu}|^2/ (\sum_{\ell} |U_{N\ell}|^2)$ for the LNV and LNC decays, respectively.}
\label{Bratio}
\end{figure}

The actual observability of these LNV and LNC decays depends on their absolute rates. This issue has been studied elsewhere \cite{Izaguirre:2015pga}, so we will  be brief here, without details on $W$ production rates or backgrounds. The $W$ branching ratios, shown in Eqs.~(\ref{bratio}) and (\ref{bratio2}),   depend on the unknown lepton mixing elements, $U_{N\ell}$ and the sterile neutrino mass $m_N$.  Since the lepton mixing is a global factor in the rate, we can divide it out to study the dependence on the neutrino mass. The branching ratios as functions of the neutrino mass $m_N$ are shown in Fig.~\ref{Bratio}. The figure exhibits the dependence on $m_N$ only, with the lepton mixing factor removed.  The actual  branching ratios are obtained  as $Br = \overline{Br}\times  |U_{Ne}|^4/ (\sum_{\ell} |U_{N\ell}|^2)$ for the LNV decay and $Br = \overline{Br}\times  |U_{Ne}U_{N\mu}|^2/ (\sum_{\ell} |U_{N\ell}|^2)$ for the LNC decay. It is clear that the rates vanish as $m_N$ approaches $M_W$, because the phase space to produce an on-shell neutrino $N$ from an on-shell $W$ vanishes as $m_N\to M_W$. At the LHC, however, the actual processes are $pp\to e^\pm e^\pm\mu^\mp\nu X$, which do not vanish but rather decrease considerably as the final state is produced through a virtual $W$ or $N$.
In this respect, these LNV and LNC lepton decays of the $W$ boson are more difficult to detect for  larger neutrino masses (still in the range below $M_W$). One should also consider that, for $m_N \lesssim 20$ GeV, even when the rates are higher according to Fig.~\ref{Bratio}, the sensitivity at the LHC is affected by the required cuts on the lepton energies.

Let us make a rough estimate of the sensitivity of the LHC to these processes. The high luminosity runs, with 3000 fb$^{-1}$, will produce about $10^{11}$ W bosons \cite{Mangano:2014xta, Grossmann:2015lea}. For the $|U_{N\ell}|^2$ mixings, according to a recent analysis \cite{Antusch:2015mia}, current representative values for their upper bounds can be taken to be $|U_{Ne}|^2 \sim 10^{-5}$ and 
$|U_{N\mu}|^2 \sim 10^{-6}$, for a neutrino mass $m_N \sim 50$ GeV. For the reduced branching ratio, as shown in  Fig.~\ref{Bratio}, we can use $\overline{Br} \sim 10^{-3}$. Consequently, for the 
 LNV mode $W^+\to  e^+ e^+ \mu^-\bar\nu_\mu$  (proportional to $|U_{Ne}|^4/ \sum |U_{N\ell}|^2 \sim 10^{-5}$)  one should get $\sim 10^3$ events,
for the LNC modes $W^+\to  e^+ e^+ \mu^-\nu_e$ or $W^+\to  \mu^+ \mu^+ e^-\nu_\mu$ 
(proportional to $|U_{Ne}|^2|U_{N\mu}|^2/ \sum |U_{N\ell}|^2 \sim 10^{-6}$)  about $ 10^2$ events,
and for the LNV mode $W^+\to  \mu^+ \mu^+ e^-\bar\nu_e$ (proportional to $|U_{N\mu}|^4/ \sum |U_{N\ell}|^2 \sim 10^{-7}$) about$10^1$ events. Assuming one can eliminate the background without getting rid of the signals, with these samples the experiment should be able to distinguish the Majorana vs. Dirac character of the sterile neutrino using the spectra of $W\to e e \mu \nu$. If the actual mixings are below these current upper bounds, the LHC will correspondingly lose sensitivity to these modes.

\begin{figure}[h]
\includegraphics[scale=0.8]{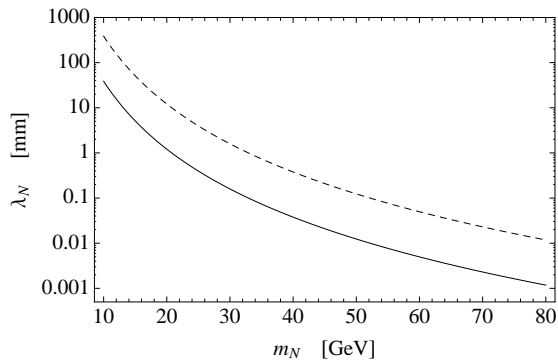}
\caption{Decay length of a heavy sterile neutrino $vs.$ its mass, $m_N$, for a Lorentz factor $\gamma=1$ and a mixing parameter $|U_{N\ell}|^2$ at its current upper bound  of $10^{-7}$ (solid line), and at lower value $10^{-9}$ (dashed line).  For other values of the parameters, in general $\lambda_N$ grows proportional to $\gamma /|U_{N\ell}|^2$. For the expressions see Eq.~(\ref{dlength}) and the Appendix. }
\label{decaylength}
\end{figure}

In addition, one should take into account that the $W$ decays we are studying may have an observable vertex separation due to the finite lifetime of the intermediate neutrino. For a neutrino $N$ with a decay width $\Gamma_N$ [see Eq.~(\ref{Nwidth})], we define its decay length $\lambda_N$ as
\begin{equation}
\lambda_N =  \gamma \frac{\hbar c}{\Gamma_N}.
\label{dlength}
\end{equation}
where $\gamma = E_N/m_N$ is the relativistic Lorentz factor. This decay length represents the typical distance between the vertex where $N$ is produced (accompanied by the primary charged lepton) and the vertex where it decays (into two charged leptons and a light neutrino). The derivation of the width as a function of the heavy neutrino mass and mixing for $m_N \gtrsim 5$ GeV is given in the Appendix. The vertex displacement is very sensitive to the neutrino mass  and the mixing parameters $|U_{N\ell}|^2$ , ($\lambda_N\sim 1/(m_N^5 |U_{N\ell}|^2)$).  As shown in Fig~\ref{decaylength}, taking the mixing at its current upper bound
$|U_{N\ell}^2 = 10^{-7}$, and without considering the relativistic factor,  the vertex separation will be above 100 $\mu m$ if  $m_N$ is below $\sim 30$ GeV. For smaller mixing values the vertex separation will be larger.

The vertex separation can be used as a condition to eliminate backgrounds. Larger separations occur for lower $m_N$. Such cases also have the additional advantage that the branching ratios are potentially larger, as discussed above. On the other hand, larger vertex separations are also obtained if the lepton mixing is smaller than the current upper bound, in which case the branching ratios become smaller, and so less likely to be detected.


%


\section{Summary}

We have studied the leptonic decays of a $W$ boson $W^+\to e^+ e^+ \mu^-\bar\nu_\mu$
and $W^+\to e^+ \mu^- e^+ \nu_e$ in order to assess their potential to discover a heavy sterile neutrino with mass below $M_W$ and, most of all, to discern between the Dirac or Majorana character of such neutrino. The first mode is a lepton number violating process (LNV) while the second preserves lepton number (LNC) but still violates lepton flavor. Both processes are mediated by an intermediate sterile neutrino. Being sterile the neutrino, the rates are suppressed by heavy-to-light lepton mixing matrix elements, $U_{N\ell}$, but on the other hand they are resonantly enhanced if the neutrino mass is in a range below $M_W$, such that it goes on its mass shell in the intermediate state. This enhancement could make possible the observation of these modes at the LHC, provided the lepton mixing elements are not too far below their current upper bounds.

Now, if the neutrino is Majorana, both LNV and LNC decays will be induced, with a relative proportion that depends on the ratio of mixings $|U_{Ne}|^2/|U_{N\mu}|^2$, but if the neutrino is Dirac, only the LNC process will occur. Therefore, a way to distinguish the Dirac $vs.$ Majorana character of the heavy neutrino is to observe which of these two processes occur.

Here we face two difficulties. On the one hand, in an actual experiment, the LNC and LNV processes cannot be distinguished by their final particles, because the only different one is the neutrino (a muon or electron neutrino), which goes undetected. On the other hand, the LNC and LNV processes cannot be distinguished by their rate, because they have exactly the same expression in terms of $m_N$, differing only in the mixing elements, which is still to date an unknown global factor. However, we find that the two processes can be distinguished by the energy spectrum of the muon ($i.e.$ the opposite sign lepton). In the LNC process, the spectrum rises continuously with energy, all the way to the endpoint, while in the LNV process, the spectrum reaches a maximum at an intermediate energy and then gradually drops to zero at the endpoint.

An additional difficulty arises due to the fact that, while a Dirac neutrino induces the LNC spectrum only, the Majorana neutrino induces both LNC and LNV processes, in relative proportion to the unknown ratio of the mixings $|U_{Ne}|^2/|U_{N\mu}|^2$, so that the shape of the spectrum induced by a Majorana neutrino is not known \emph{a priori}. We therefore have studied the difference in the spectra as a function of this ratio of mixings. Due to the fact the LNV process is proportional to $|U_{Ne}|^4$ while the LNC process is proportional to $|U_{N e} U_{N\mu}|^2$,  it turns out that the observable LNV spectrum is more distinguished from the LNC spectrum as  $|U_{N e} /U_{N\mu}|^2$ is larger, and become confused as this ratio becomes  less than unity. Consequently the separation between Dirac $vs.$ Majorana cannot be done with these processes if this mixing ratio turns out to be small. However, in that case one should use the 
decays $W^+\to \mu^+\mu^+ e^- \bar\nu_e$ and $W^+\to \mu^+ e^- \mu^+ \nu_\mu$, where the muon and electron roles are exchanged. With these processes, the distinction between the LNC and LNV spectra becomes clearer as the ratio $|U_{N e} /U_{N\mu}|^2$ gets smaller, 
precisely where the original processes become confused, in a complementary way.

\acknowledgments

C.D. thanks the hospitality of the ICTP-SAIFR, Sao Paulo, Brazil, with the support of FAPESP grant 2011/11973-4, where part of this work was done. The work of C.D. was supported  in part by FONDECYT Grant No.~1130617. The work of C.S.K. was supported by the NRF grant funded by the Korean government of the MEST (No. 2011-0017430) and (No. 2011-0020333).

\appendix

\section{The heavy neutrino width}


The following is an estimate of the heavy neutrino width, formulated consistently within a generic seesaw scenario. For $m_N \gtrsim 5$ GeV, we approximate the width by considering just open quark and lepton channels, and neglecting fermion masses of the final states whenever the corresponding channel is kinematically open.
As shown in Atre, Han \& Pascoli, the channels are:

1) Charged currents in leptonic decay with $\ell_1\neq \ell_2$:
%
\begin{equation}
\Gamma(N\to \ell_1^- \ell_2^+ \nu_{\ell_2}) = \frac{G_F^2}{192 \pi^3}m_N^5 |U_{N\ell_1}|^2 .
\label{chargelepton}
\end{equation}
For each $\ell_1$ there are two $\ell_2$ flavours, and then the conjugate channels give an extra factor 2. In total we get a factor 4 for each $\ell_1$.


2) Neutral currents into a charged lepton pair with $\ell_1\neq \ell_2$:
\begin{equation}
\Gamma(N\to \nu_1 \ell_2^+\ell_2^-) = \frac{G_F^2}{96 \pi^3}m_N^5 |U_{N\ell_1}|^2
\times \Big\{
g_L g_R  + g_L^2 + g_R^2
\Big\},
\label{neutralchargelepton}
\end{equation}
where the chiral couplings to $Z$ are
%
%
$g_L = - \frac{1}{2} +   \sin^2\theta_W $
and $g_R =  \sin^2\theta_W $.  For a given $\ell_1$, there are 2 possible $\ell_2$ flavours:
we get a factor 2 for each $\ell_1$.


3) Charged and neutral currents for the case $\ell_1 = \ell_2$:
\begin{equation}
\Gamma(N\to \nu_1 \ell_1^+\ell_1^-) = \frac{G_F^2}{96 \pi^3}m_N^5 |U_{N\ell_1}|^2
\times \Big\{
g_L g_R  + g_L^2 + g_R^2 + 1+2 g_L+ g_R
\Big\},
\label{neutralequallepton}
\end{equation}
where $g_L = -\frac{1}{2} + \sin^2 \theta_W$ and $g_R = \sin^2 \theta_W$.
For a given  $\ell_1$, there is only one case.


4) Neutral currents into neutrinos:
\begin{equation}
\Gamma(N\to \nu_1 \nu_2\bar \nu_2) = \frac{G_F^2}{768 \pi^3}m_N^5 |U_{N\ell_1}|^2 .
\end{equation}
Considering all $\nu_2$ flavors including $\nu_2 = \nu_1$, the sum over all $\nu_2$ gives factor 3 for each $\nu_1$.
%


5) Purely charged currents with quarks.  We neglect quark masses.
The final fermion pairs are $f_u \bar f_d= $ $u\bar d$ and $c \bar s$, and we  must include charged conjugates as well:
\begin{equation}
\Gamma(N\to \ell_1^- f_u \bar f_d) = \frac{G_F^2}{64 \pi^3}m_N^5 |U_{N\ell_1}|^2 .
\label{chargequark}
\end{equation}
This expression differs from the lepton case just by a factor 3 due to color. There are 2 quark channels, and then the conjugate channels give an extra factor 2. In total we get a factor 4 for each $\ell_1$.

6) Neutral currents into same flavour quark-antiquark ($q\bar q$, with $q= u, d, c, s, b$):
\begin{equation}
\Gamma(N\to \nu_1 q \bar q) = \frac{G_F^2}{32 \pi^3}m_N^5 |U_{N\ell_1}|^2
\times \Big\{
g_L g_R  + g_L^2 + g_R^2
\Big\}.
\label{neutralquark}
\end{equation}
For up-type quarks, $g_L = \frac{1}{2} - \frac{2}{3}  \sin^2\theta_W $
and $g_R = - \frac{2}{3} \sin^2\theta_W $. Summing over  the two up-type flavours, we get a factor 2 for each $\ell_1$.
For down-type quarks,
$g_L =-\frac{1}{2} + \frac{1}{3}  \sin^2\theta_W $ and $g_R = \frac{1}{3}  \sin^2\theta_W $.
There are three  down-type flavours, so we get a factor 3 for each $\ell_1$.


The total width is then:
\begin{eqnarray}
\Gamma_N = \sum_{\ell_1} &&
4\ \Gamma(N\to \ell_1^- \ell_2^+ \nu_{\ell_2})
+
2\  \Gamma(N\to \nu_1 \ell_2^+ \ell_2^-)
+
 \Gamma(N\to \nu_1 \ell_1^+\ell_1^-)
 +
 3\   \Gamma(N\to \nu_1 \nu_2\bar\nu_2) \nonumber \\
&& + \
4 \ \Gamma(N\to \ell_1^- f_u \bar f_d)
+
2\  \Gamma(N\to \nu_1 f_u \bar f_u)
+
3 \ \Gamma(N\to \nu_1 f_d \bar f_d)
,
\end{eqnarray}
\begin{eqnarray}
\Gamma_N  \approx 0.116 \times \frac{G_F^2}{\pi^3} m_N^5 \sum_{\ell_1} |U_{N\ell_1}|^2  .
\label{Nwidth}
\end{eqnarray}


%
%
%
%
%

\bibliographystyle{apsrev4-1}
\bibliography{Ref}

\begin{thebibliography}{38}%
\makeatletter
\providecommand \@ifxundefined [1]{%
 \@ifx{#1\undefined}
}%
\providecommand \@ifnum [1]{%
 \ifnum #1\expandafter \@firstoftwo
 \else \expandafter \@secondoftwo
 \fi
}%
\providecommand \@ifx [1]{%
 \ifx #1\expandafter \@firstoftwo
 \else \expandafter \@secondoftwo
 \fi
}%
\providecommand \natexlab [1]{#1}%
\providecommand \enquote  [1]{``#1''}%
\providecommand \bibnamefont  [1]{#1}%
\providecommand \bibfnamefont [1]{#1}%
\providecommand \citenamefont [1]{#1}%
\providecommand \href@noop [0]{\@secondoftwo}%
\providecommand \href [0]{\begingroup \@sanitize@url \@href}%
\providecommand \@href[1]{\@@startlink{#1}\@@href}%
\providecommand \@@href[1]{\endgroup#1\@@endlink}%
\providecommand \@sanitize@url [0]{\catcode `\\12\catcode `\$12\catcode
  `\&12\catcode `\#12\catcode `\^12\catcode `\_12\catcode `\%12\relax}%
\providecommand \@@startlink[1]{}%
\providecommand \@@endlink[0]{}%
\providecommand \url  [0]{\begingroup\@sanitize@url \@url }%
\providecommand \@url [1]{\endgroup\@href {#1}{\urlprefix }}%
\providecommand \urlprefix  [0]{URL }%
\providecommand \Eprint [0]{\href }%
\providecommand \doibase [0]{http://dx.doi.org/}%
\providecommand \selectlanguage [0]{\@gobble}%
\providecommand \bibinfo  [0]{\@secondoftwo}%
\providecommand \bibfield  [0]{\@secondoftwo}%
\providecommand \translation [1]{[#1]}%
\providecommand \BibitemOpen [0]{}%
\providecommand \bibitemStop [0]{}%
\providecommand \bibitemNoStop [0]{.\EOS\space}%
\providecommand \EOS [0]{\spacefactor3000\relax}%
\providecommand \BibitemShut  [1]{\csname bibitem#1\endcsname}%
\let\auto@bib@innerbib\@empty
\bibitem [{\citenamefont {Olive}\ \emph {et~al.}(2014)\citenamefont {Olive}
  \emph {et~al.}}]{Agashe:2014kda}%
  \BibitemOpen
  \bibfield  {author} {\bibinfo {author} {\bibfnamefont {K.~A.}\ \bibnamefont
  {Olive}} \emph {et~al.} (\bibinfo {collaboration} {Particle Data Group}),\
  }\href {\doibase 10.1088/1674-1137/38/9/090001} {\bibfield  {journal}
  {\bibinfo  {journal} {Chin. Phys.}\ }\textbf {\bibinfo {volume} {C38}},\
  \bibinfo {pages} {090001} (\bibinfo {year} {2014})}\BibitemShut {NoStop}%
\bibitem [{\citenamefont {Minkowski}(1977)}]{Minkowski:1977sc}%
  \BibitemOpen
  \bibfield  {author} {\bibinfo {author} {\bibfnamefont {P.}~\bibnamefont
  {Minkowski}},\ }\href {\doibase 10.1016/0370-2693(77)90435-X} {\bibfield
  {journal} {\bibinfo  {journal} {Phys. Lett. B}\ }\textbf {\bibinfo {volume}
  {67}},\ \bibinfo {pages} {421} (\bibinfo {year} {1977})}\BibitemShut
  {NoStop}%
\bibitem [{\citenamefont {Gell-Mann}\ \emph {et~al.}(1979)\citenamefont
  {Gell-Mann}, \citenamefont {Ramond},\ and\ \citenamefont
  {Slansky}}]{GellMann:1980vs}%
  \BibitemOpen
  \bibfield  {author} {\bibinfo {author} {\bibfnamefont {M.}~\bibnamefont
  {Gell-Mann}}, \bibinfo {author} {\bibfnamefont {P.}~\bibnamefont {Ramond}}, \
  and\ \bibinfo {author} {\bibfnamefont {R.}~\bibnamefont {Slansky}},\
  }\bibfield  {booktitle} {\emph {\bibinfo {booktitle} {{Supergravity Workshop
  Stony Brook, New York, September 27-28, 1979}}},\ }\href@noop {} {\bibfield
  {journal} {\bibinfo  {journal} {Conf. Proc.}\ }\textbf {\bibinfo {volume}
  {C790927}},\ \bibinfo {pages} {315} (\bibinfo {year} {1979})},\ \Eprint
  {http://arxiv.org/abs/1306.4669} {arXiv:1306.4669 [hep-th]} \BibitemShut
  {NoStop}%
\bibitem [{\citenamefont {Yanagida}(1979)}]{Yanagida:1979as}%
  \BibitemOpen
  \bibfield  {author} {\bibinfo {author} {\bibfnamefont {T.}~\bibnamefont
  {Yanagida}},\ }\bibfield  {booktitle} {\emph {\bibinfo {booktitle}
  {{Proceedings: Workshop on the Unified Theories and the Baryon Number in the
  Universe, Tsukuba, Japan, 13-14 Feb 1979}}},\ }\href@noop {} {\bibfield
  {journal} {\bibinfo  {journal} {Conf. Proc.}\ }\textbf {\bibinfo {volume}
  {C7902131}},\ \bibinfo {pages} {95} (\bibinfo {year} {1979})},\ \bibinfo
  {note} {[Conf. Proc.C7902131,95(1979)]}\BibitemShut {NoStop}%
\bibitem [{\citenamefont {Mohapatra}\ and\ \citenamefont
  {Senjanovi\'{c}}(1980)}]{Mohapatra:1979ia}%
  \BibitemOpen
  \bibfield  {author} {\bibinfo {author} {\bibfnamefont {R.~N.}\ \bibnamefont
  {Mohapatra}}\ and\ \bibinfo {author} {\bibfnamefont {G.}~\bibnamefont
  {Senjanovi\'{c}}},\ }\href {\doibase 10.1103/PhysRevLett.44.912} {\bibfield
  {journal} {\bibinfo  {journal} {Phys. Rev. Lett.}\ }\textbf {\bibinfo
  {volume} {44}},\ \bibinfo {pages} {912} (\bibinfo {year} {1980})}\BibitemShut
  {NoStop}%
\bibitem [{\citenamefont {Glashow}(1980)}]{Glashow:1979nm}%
  \BibitemOpen
  \bibfield  {author} {\bibinfo {author} {\bibfnamefont {S.~L.}\ \bibnamefont
  {Glashow}},\ }\bibfield  {booktitle} {\emph {\bibinfo {booktitle} {{Cargese
  Summer Institute: Recent Developments in Gauge Theories Cargese, France,
  August 26-September 8, 1979}}},\ }\href {\doibase
  10.1007/978-1-4684-7197-7_15} {\bibfield  {journal} {\bibinfo  {journal}
  {NATO Sci. Ser. B}\ }\textbf {\bibinfo {volume} {61}},\ \bibinfo {pages}
  {687} (\bibinfo {year} {1980})}\BibitemShut {NoStop}%
\bibitem [{\citenamefont {Schechter}\ and\ \citenamefont
  {Valle}(1980)}]{Schechter:1980gr}%
  \BibitemOpen
  \bibfield  {author} {\bibinfo {author} {\bibfnamefont {J.}~\bibnamefont
  {Schechter}}\ and\ \bibinfo {author} {\bibfnamefont {J.~W.~F.}\ \bibnamefont
  {Valle}},\ }\href {\doibase 10.1103/PhysRevD.22.2227} {\bibfield  {journal}
  {\bibinfo  {journal} {Phys. Rev. D}\ }\textbf {\bibinfo {volume} {22}},\
  \bibinfo {pages} {2227} (\bibinfo {year} {1980})}\BibitemShut {NoStop}%
\bibitem [{\citenamefont {Ma}(2009)}]{Ma:2009dk}%
  \BibitemOpen
  \bibfield  {author} {\bibinfo {author} {\bibfnamefont {E.}~\bibnamefont
  {Ma}},\ }\href@noop {} {\  (\bibinfo {year} {2009})},\ \Eprint
  {http://arxiv.org/abs/0905.0221} {arXiv:0905.0221 [hep-ph]} \BibitemShut
  {NoStop}%
\bibitem [{\citenamefont {Fileviez~Perez}\ and\ \citenamefont
  {Wise}(2009)}]{FileviezPerez:2009ud}%
  \BibitemOpen
  \bibfield  {author} {\bibinfo {author} {\bibfnamefont {P.}~\bibnamefont
  {Fileviez~Perez}}\ and\ \bibinfo {author} {\bibfnamefont {M.~B.}\
  \bibnamefont {Wise}},\ }\href {\doibase 10.1103/PhysRevD.80.053006}
  {\bibfield  {journal} {\bibinfo  {journal} {Phys. Rev.}\ }\textbf {\bibinfo
  {volume} {D80}},\ \bibinfo {pages} {053006} (\bibinfo {year} {2009})},\
  \Eprint {http://arxiv.org/abs/0906.2950} {arXiv:0906.2950 [hep-ph]}
  \BibitemShut {NoStop}%
\bibitem [{\citenamefont {Keung}\ and\ \citenamefont
  {Senjanovi\'{c}}(1983)}]{Keung:1983uu}%
  \BibitemOpen
  \bibfield  {author} {\bibinfo {author} {\bibfnamefont {W.-Y.}\ \bibnamefont
  {Keung}}\ and\ \bibinfo {author} {\bibfnamefont {G.}~\bibnamefont
  {Senjanovi\'{c}}},\ }\href {\doibase 10.1103/PhysRevLett.50.1427} {\bibfield
  {journal} {\bibinfo  {journal} {Phys. Rev. Lett.}\ }\textbf {\bibinfo
  {volume} {50}},\ \bibinfo {pages} {1427} (\bibinfo {year}
  {1983})}\BibitemShut {NoStop}%
\bibitem [{\citenamefont {Atre}\ \emph {et~al.}(2009)\citenamefont {Atre},
  \citenamefont {Han}, \citenamefont {Pascoli},\ and\ \citenamefont
  {Zhang}}]{Atre:2009rg}%
  \BibitemOpen
  \bibfield  {author} {\bibinfo {author} {\bibfnamefont {A.}~\bibnamefont
  {Atre}}, \bibinfo {author} {\bibfnamefont {T.}~\bibnamefont {Han}}, \bibinfo
  {author} {\bibfnamefont {S.}~\bibnamefont {Pascoli}}, \ and\ \bibinfo
  {author} {\bibfnamefont {B.}~\bibnamefont {Zhang}},\ }\href {\doibase
  10.1088/1126-6708/2009/05/030} {\bibfield  {journal} {\bibinfo  {journal}
  {JHEP}\ }\textbf {\bibinfo {volume} {0905}},\ \bibinfo {pages} {030}
  (\bibinfo {year} {2009})},\ \Eprint {http://arxiv.org/abs/0901.3589}
  {arXiv:0901.3589 [hep-ph]} \BibitemShut {NoStop}%
\bibitem [{\citenamefont {Maneschg}(2015)}]{Maneschg:2015dja}%
  \BibitemOpen
  \bibfield  {author} {\bibinfo {author} {\bibfnamefont {W.}~\bibnamefont
  {Maneschg}},\ }\bibfield  {booktitle} {\emph {\bibinfo {booktitle}
  {{Proceedings, 13th International Workshop on Tau Lepton Physics (TAU
  2014)}}},\ }\href {\doibase 10.1016/j.nuclphysbps.2015.02.039} {\bibfield
  {journal} {\bibinfo  {journal} {Nucl. Part. Phys. Proc.}\ }\textbf {\bibinfo
  {volume} {260}},\ \bibinfo {pages} {188} (\bibinfo {year}
  {2015})}\BibitemShut {NoStop}%
\bibitem [{\citenamefont {Simkovic}(2013)}]{Simkovic:2013kna}%
  \BibitemOpen
  \bibfield  {author} {\bibinfo {author} {\bibfnamefont {F.}~\bibnamefont
  {Simkovic}},\ }\href {\doibase 10.1134/S154747711307008X} {\bibfield
  {journal} {\bibinfo  {journal} {Phys. Part. Nucl. Lett.}\ }\textbf {\bibinfo
  {volume} {10}},\ \bibinfo {pages} {623} (\bibinfo {year} {2013})}\BibitemShut
  {NoStop}%
\bibitem [{\citenamefont {Bilenky}\ and\ \citenamefont
  {Giunti}(2012)}]{Bilenky:2012qi}%
  \BibitemOpen
  \bibfield  {author} {\bibinfo {author} {\bibfnamefont {S.~M.}\ \bibnamefont
  {Bilenky}}\ and\ \bibinfo {author} {\bibfnamefont {C.}~\bibnamefont
  {Giunti}},\ }\href {\doibase 10.1142/S0217732312300157} {\bibfield  {journal}
  {\bibinfo  {journal} {Mod. Phys. Lett.}\ }\textbf {\bibinfo {volume} {A27}},\
  \bibinfo {pages} {1230015} (\bibinfo {year} {2012})},\ \Eprint
  {http://arxiv.org/abs/1203.5250} {arXiv:1203.5250 [hep-ph]} \BibitemShut
  {NoStop}%
\bibitem [{\citenamefont {Faessler}(2009)}]{Faessler:2009zz}%
  \BibitemOpen
  \bibfield  {author} {\bibinfo {author} {\bibfnamefont {A.}~\bibnamefont
  {Faessler}},\ }\bibfield  {booktitle} {\emph {\bibinfo {booktitle} {{NOW
  2008, proceedings of the Neutrino Oscillation Workshop, Conca Specchiulla,
  Otranto, Italy, 6-12 September 2008}}},\ }\href {\doibase
  10.1016/j.nuclphysbps.2009.02.003} {\bibfield  {journal} {\bibinfo  {journal}
  {Nucl. Phys. Proc. Suppl.}\ }\textbf {\bibinfo {volume} {188}},\ \bibinfo
  {pages} {20} (\bibinfo {year} {2009})}\BibitemShut {NoStop}%
\bibitem [{\citenamefont {Cvetic}\ \emph {et~al.}(2010)\citenamefont {Cvetic},
  \citenamefont {Dib}, \citenamefont {Kang},\ and\ \citenamefont
  {Kim}}]{Cvetic:2010rw}%
  \BibitemOpen
  \bibfield  {author} {\bibinfo {author} {\bibfnamefont {G.}~\bibnamefont
  {Cvetic}}, \bibinfo {author} {\bibfnamefont {C.~O.}\ \bibnamefont {Dib}},
  \bibinfo {author} {\bibfnamefont {S.~K.}\ \bibnamefont {Kang}}, \ and\
  \bibinfo {author} {\bibfnamefont {C.~S.}\ \bibnamefont {Kim}},\ }\href
  {\doibase 10.1103/PhysRevD.82.053010} {\bibfield  {journal} {\bibinfo
  {journal} {Phys. Rev. D}\ }\textbf {\bibinfo {volume} {82}},\ \bibinfo
  {pages} {053010} (\bibinfo {year} {2010})},\ \Eprint
  {http://arxiv.org/abs/1005.4282} {arXiv:1005.4282 [hep-ph]} \BibitemShut
  {NoStop}%
\bibitem [{\citenamefont {Cvetic}\ \emph {et~al.}(2012)\citenamefont {Cvetic},
  \citenamefont {Dib},\ and\ \citenamefont {Kim}}]{Cvetic:2012hd}%
  \BibitemOpen
  \bibfield  {author} {\bibinfo {author} {\bibfnamefont {G.}~\bibnamefont
  {Cvetic}}, \bibinfo {author} {\bibfnamefont {C.~O.}\ \bibnamefont {Dib}}, \
  and\ \bibinfo {author} {\bibfnamefont {C.~S.}\ \bibnamefont {Kim}},\ }\href
  {\doibase 10.1007/JHEP06(2012)149} {\bibfield  {journal} {\bibinfo  {journal}
  {JHEP}\ }\textbf {\bibinfo {volume} {1206}},\ \bibinfo {pages} {149}
  (\bibinfo {year} {2012})},\ \Eprint {http://arxiv.org/abs/1203.0573}
  {arXiv:1203.0573 [hep-ph]} \BibitemShut {NoStop}%
\bibitem [{\citenamefont {Lees}\ \emph {et~al.}(2014)\citenamefont {Lees} \emph
  {et~al.}}]{Lees:2013gdj}%
  \BibitemOpen
  \bibfield  {author} {\bibinfo {author} {\bibfnamefont {J.}~\bibnamefont
  {Lees}} \emph {et~al.} (\bibinfo {collaboration} {BaBar}),\ }\href {\doibase
  10.1103/PhysRevD.89.011102} {\bibfield  {journal} {\bibinfo  {journal} {Phys.
  Rev. D}\ }\textbf {\bibinfo {volume} {89}},\ \bibinfo {pages} {011102}
  (\bibinfo {year} {2014})},\ \Eprint {http://arxiv.org/abs/1310.8238}
  {arXiv:1310.8238 [hep-ex]} \BibitemShut {NoStop}%
\bibitem [{\citenamefont {Liventsev}\ \emph {et~al.}(2013)\citenamefont
  {Liventsev} \emph {et~al.}}]{Liventsev:2013zz}%
  \BibitemOpen
  \bibfield  {author} {\bibinfo {author} {\bibfnamefont {D.}~\bibnamefont
  {Liventsev}} \emph {et~al.} (\bibinfo {collaboration} {Belle}),\ }\href
  {\doibase 10.1103/PhysRevD.87.071102} {\bibfield  {journal} {\bibinfo
  {journal} {Phys. Rev. D}\ }\textbf {\bibinfo {volume} {87}},\ \bibinfo
  {pages} {071102} (\bibinfo {year} {2013})},\ \Eprint
  {http://arxiv.org/abs/1301.1105} {arXiv:1301.1105 [hep-ex]} \BibitemShut
  {NoStop}%
\bibitem [{\citenamefont {Bonivento}\ \emph {et~al.}(2013)\citenamefont
  {Bonivento} \emph {et~al.}}]{Bonivento:2013jag}%
  \BibitemOpen
  \bibfield  {author} {\bibinfo {author} {\bibfnamefont {W.}~\bibnamefont
  {Bonivento}} \emph {et~al.},\ }\href@noop {} {\  (\bibinfo {year} {2013})},\
  \Eprint {http://arxiv.org/abs/1310.1762} {arXiv:1310.1762 [hep-ex]}
  \BibitemShut {NoStop}%
\bibitem [{\citenamefont {Dib}\ and\ \citenamefont {Kim}(2014)}]{Dib:2014iga}%
  \BibitemOpen
  \bibfield  {author} {\bibinfo {author} {\bibfnamefont {C.~O.}\ \bibnamefont
  {Dib}}\ and\ \bibinfo {author} {\bibfnamefont {C.~S.}\ \bibnamefont {Kim}},\
  }\href {\doibase 10.1103/PhysRevD.89.077301} {\bibfield  {journal} {\bibinfo
  {journal} {Phys. Rev. D}\ }\textbf {\bibinfo {volume} {89}},\ \bibinfo
  {pages} {077301} (\bibinfo {year} {2014})},\ \Eprint
  {http://arxiv.org/abs/1403.1985} {arXiv:1403.1985 [hep-ph]} \BibitemShut
  {NoStop}%
\bibitem [{\citenamefont {Deppisch}\ \emph {et~al.}(2015)\citenamefont
  {Deppisch}, \citenamefont {Dev},\ and\ \citenamefont
  {Pilaftsis}}]{Deppisch:2015qwa}%
  \BibitemOpen
  \bibfield  {author} {\bibinfo {author} {\bibfnamefont {F.~F.}\ \bibnamefont
  {Deppisch}}, \bibinfo {author} {\bibfnamefont {P.~S.~B.}\ \bibnamefont
  {Dev}}, \ and\ \bibinfo {author} {\bibfnamefont {A.}~\bibnamefont
  {Pilaftsis}},\ }\href@noop {} {\  (\bibinfo {year} {2015})},\ \Eprint
  {http://arxiv.org/abs/1502.06541} {arXiv:1502.06541 [hep-ph]} \BibitemShut
  {NoStop}%
\bibitem [{\citenamefont {Abada}(2013)}]{Abada:2013bpa}%
  \BibitemOpen
  \bibfield  {author} {\bibinfo {author} {\bibfnamefont {A.}~\bibnamefont
  {Abada}},\ }in\ \href
  {http://inspirehep.net/record/1258427/files/arXiv:1310.3800.pdf} {\emph
  {\bibinfo {booktitle} {{25th Rencontres de Blois on Particle Physics and
  Cosmology Blois, France, May 26-31, 2013}}}}\ (\bibinfo {year} {2013})\
  \Eprint {http://arxiv.org/abs/1310.3800} {arXiv:1310.3800 [hep-ph]}
  \BibitemShut {NoStop}%
\bibitem [{\citenamefont {Das}\ and\ \citenamefont {Okada}(2013)}]{Das:2012ze}%
  \BibitemOpen
  \bibfield  {author} {\bibinfo {author} {\bibfnamefont {A.}~\bibnamefont
  {Das}}\ and\ \bibinfo {author} {\bibfnamefont {N.}~\bibnamefont {Okada}},\
  }\href {\doibase 10.1103/PhysRevD.88.113001} {\bibfield  {journal} {\bibinfo
  {journal} {Phys. Rev.}\ }\textbf {\bibinfo {volume} {D88}},\ \bibinfo {pages}
  {113001} (\bibinfo {year} {2013})},\ \Eprint {http://arxiv.org/abs/1207.3734}
  {arXiv:1207.3734 [hep-ph]} \BibitemShut {NoStop}%
\bibitem [{\citenamefont {Das}\ \emph {et~al.}(2014)\citenamefont {Das},
  \citenamefont {Dev},\ and\ \citenamefont {Okada}}]{Das:2014jxa}%
  \BibitemOpen
  \bibfield  {author} {\bibinfo {author} {\bibfnamefont {A.}~\bibnamefont
  {Das}}, \bibinfo {author} {\bibfnamefont {P.}~\bibnamefont {Dev}}, \ and\
  \bibinfo {author} {\bibfnamefont {N.}~\bibnamefont {Okada}},\ }\href
  {\doibase 10.1016/j.physletb.2014.06.058} {\bibfield  {journal} {\bibinfo
  {journal} {Phys. Lett.}\ }\textbf {\bibinfo {volume} {B735}},\ \bibinfo
  {pages} {364} (\bibinfo {year} {2014})},\ \Eprint
  {http://arxiv.org/abs/1405.0177} {arXiv:1405.0177 [hep-ph]} \BibitemShut
  {NoStop}%
\bibitem [{\citenamefont {Bar-Shalom}\ \emph {et~al.}(2006)\citenamefont
  {Bar-Shalom}, \citenamefont {Deshpande}, \citenamefont {Eilam}, \citenamefont
  {Jiang},\ and\ \citenamefont {Soni}}]{BarShalom:2006bv}%
  \BibitemOpen
  \bibfield  {author} {\bibinfo {author} {\bibfnamefont {S.}~\bibnamefont
  {Bar-Shalom}}, \bibinfo {author} {\bibfnamefont {N.~G.}\ \bibnamefont
  {Deshpande}}, \bibinfo {author} {\bibfnamefont {G.}~\bibnamefont {Eilam}},
  \bibinfo {author} {\bibfnamefont {J.}~\bibnamefont {Jiang}}, \ and\ \bibinfo
  {author} {\bibfnamefont {A.}~\bibnamefont {Soni}},\ }\href {\doibase
  10.1016/j.physletb.2006.10.060} {\bibfield  {journal} {\bibinfo  {journal}
  {Phys. Lett.}\ }\textbf {\bibinfo {volume} {B643}},\ \bibinfo {pages} {342}
  (\bibinfo {year} {2006})},\ \Eprint {http://arxiv.org/abs/hep-ph/0608309}
  {arXiv:hep-ph/0608309 [hep-ph]} \BibitemShut {NoStop}%
\bibitem [{\citenamefont {del Aguila}\ and\ \citenamefont
  {Aguilar-Saavedra}(2009)}]{delAguila:2008cj}%
  \BibitemOpen
  \bibfield  {author} {\bibinfo {author} {\bibfnamefont {F.}~\bibnamefont {del
  Aguila}}\ and\ \bibinfo {author} {\bibfnamefont {J.~A.}\ \bibnamefont
  {Aguilar-Saavedra}},\ }\href {\doibase 10.1016/j.nuclphysb.2008.12.029}
  {\bibfield  {journal} {\bibinfo  {journal} {Nucl. Phys.}\ }\textbf {\bibinfo
  {volume} {B813}},\ \bibinfo {pages} {22} (\bibinfo {year} {2009})},\ \Eprint
  {http://arxiv.org/abs/0808.2468} {arXiv:0808.2468 [hep-ph]} \BibitemShut
  {NoStop}%
\bibitem [{\citenamefont {Aad}\ \emph {et~al.}(2015)\citenamefont {Aad} \emph
  {et~al.}}]{Aad:2015xaa}%
  \BibitemOpen
  \bibfield  {author} {\bibinfo {author} {\bibfnamefont {G.}~\bibnamefont
  {Aad}} \emph {et~al.} (\bibinfo {collaboration} {ATLAS}),\ }\href {\doibase
  10.1007/JHEP07(2015)162} {\bibfield  {journal} {\bibinfo  {journal} {JHEP}\
  }\textbf {\bibinfo {volume} {07}},\ \bibinfo {pages} {162} (\bibinfo {year}
  {2015})},\ \Eprint {http://arxiv.org/abs/1506.06020} {arXiv:1506.06020
  [hep-ex]} \BibitemShut {NoStop}%
\bibitem [{\citenamefont {Khachatryan}\ \emph {et~al.}(2015)\citenamefont
  {Khachatryan} \emph {et~al.}}]{Khachatryan:2015gha}%
  \BibitemOpen
  \bibfield  {author} {\bibinfo {author} {\bibfnamefont {V.}~\bibnamefont
  {Khachatryan}} \emph {et~al.} (\bibinfo {collaboration} {CMS}),\ }\href
  {\doibase 10.1016/j.physletb.2015.06.070} {\bibfield  {journal} {\bibinfo
  {journal} {Phys. Lett.}\ }\textbf {\bibinfo {volume} {B748}},\ \bibinfo
  {pages} {144} (\bibinfo {year} {2015})},\ \Eprint
  {http://arxiv.org/abs/1501.05566} {arXiv:1501.05566 [hep-ex]} \BibitemShut
  {NoStop}%
\bibitem [{\citenamefont {Helo}\ \emph {et~al.}(2014)\citenamefont {Helo},
  \citenamefont {Hirsch},\ and\ \citenamefont {Kovalenko}}]{Helo:2013esa}%
  \BibitemOpen
  \bibfield  {author} {\bibinfo {author} {\bibfnamefont {J.~C.}\ \bibnamefont
  {Helo}}, \bibinfo {author} {\bibfnamefont {M.}~\bibnamefont {Hirsch}}, \ and\
  \bibinfo {author} {\bibfnamefont {S.}~\bibnamefont {Kovalenko}},\ }\href
  {\doibase 10.1103/PhysRevD.89.073005} {\bibfield  {journal} {\bibinfo
  {journal} {Phys. Rev. D}\ }\textbf {\bibinfo {volume} {89}},\ \bibinfo
  {pages} {073005} (\bibinfo {year} {2014})},\ \Eprint
  {http://arxiv.org/abs/1312.2900} {arXiv:1312.2900 [hep-ph]} \BibitemShut
  {NoStop}%
\bibitem [{\citenamefont {Izaguirre}\ and\ \citenamefont
  {Shuve}(2015)}]{Izaguirre:2015pga}%
  \BibitemOpen
  \bibfield  {author} {\bibinfo {author} {\bibfnamefont {E.}~\bibnamefont
  {Izaguirre}}\ and\ \bibinfo {author} {\bibfnamefont {B.}~\bibnamefont
  {Shuve}},\ }\href {\doibase 10.1103/PhysRevD.91.093010} {\bibfield  {journal}
  {\bibinfo  {journal} {Phys. Rev. D}\ }\textbf {\bibinfo {volume} {91}},\
  \bibinfo {pages} {093010} (\bibinfo {year} {2015})},\ \Eprint
  {http://arxiv.org/abs/1504.02470} {arXiv:1504.02470 [hep-ph]} \BibitemShut
  {NoStop}%
\bibitem [{\citenamefont {Gago}\ \emph {et~al.}(2015)\citenamefont {Gago},
  \citenamefont {Hern{\'a}ndez}, \citenamefont {Jones-P{\'e}rez}, \citenamefont
  {Losada},\ and\ \citenamefont {Brice{\~n}o}}]{Gago:2015vma}%
  \BibitemOpen
  \bibfield  {author} {\bibinfo {author} {\bibfnamefont {A.~M.}\ \bibnamefont
  {Gago}}, \bibinfo {author} {\bibfnamefont {P.}~\bibnamefont {Hern{\'a}ndez}},
  \bibinfo {author} {\bibfnamefont {J.}~\bibnamefont {Jones-P{\'e}rez}},
  \bibinfo {author} {\bibfnamefont {M.}~\bibnamefont {Losada}}, \ and\ \bibinfo
  {author} {\bibfnamefont {A.~M.}\ \bibnamefont {Brice{\~n}o}},\ }\href@noop {}
  {\  (\bibinfo {year} {2015})},\ \Eprint {http://arxiv.org/abs/1505.05880}
  {arXiv:1505.05880 [hep-ph]} \BibitemShut {NoStop}%
\bibitem [{\citenamefont {Cortes}\ \emph {et~al.}(1986)\citenamefont {Cortes},
  \citenamefont {Hagiwara},\ and\ \citenamefont {Herzog}}]{Cortes:1985yi}%
  \BibitemOpen
  \bibfield  {author} {\bibinfo {author} {\bibfnamefont {J.}~\bibnamefont
  {Cortes}}, \bibinfo {author} {\bibfnamefont {K.}~\bibnamefont {Hagiwara}}, \
  and\ \bibinfo {author} {\bibfnamefont {F.}~\bibnamefont {Herzog}},\ }\href
  {\doibase 10.1016/0550-3213(86)90105-7} {\bibfield  {journal} {\bibinfo
  {journal} {Nucl. Phys.}\ }\textbf {\bibinfo {volume} {B278}},\ \bibinfo
  {pages} {26} (\bibinfo {year} {1986})}\BibitemShut {NoStop}%
\bibitem [{\citenamefont {Halzen}\ \emph {et~al.}(2013)\citenamefont {Halzen},
  \citenamefont {Jeong},\ and\ \citenamefont {Kim}}]{Halzen:2013bqa}%
  \BibitemOpen
  \bibfield  {author} {\bibinfo {author} {\bibfnamefont {F.}~\bibnamefont
  {Halzen}}, \bibinfo {author} {\bibfnamefont {Y.~S.}\ \bibnamefont {Jeong}}, \
  and\ \bibinfo {author} {\bibfnamefont {C.~S.}\ \bibnamefont {Kim}},\ }\href
  {\doibase 10.1103/PhysRevD.88.073013} {\bibfield  {journal} {\bibinfo
  {journal} {Phys. Rev.}\ }\textbf {\bibinfo {volume} {D88}},\ \bibinfo {pages}
  {073013} (\bibinfo {year} {2013})},\ \Eprint {http://arxiv.org/abs/1304.0322}
  {arXiv:1304.0322 [hep-ph]} \BibitemShut {NoStop}%
\bibitem [{\citenamefont {Abbott}\ \emph {et~al.}(1998)\citenamefont {Abbott}
  \emph {et~al.}}]{Abbott:1997ww}%
  \BibitemOpen
  \bibfield  {author} {\bibinfo {author} {\bibfnamefont {B.}~\bibnamefont
  {Abbott}} \emph {et~al.} (\bibinfo {collaboration} {D0}),\ }\href {\doibase
  10.1103/PhysRevD.58.092003} {\bibfield  {journal} {\bibinfo  {journal} {Phys.
  Rev.}\ }\textbf {\bibinfo {volume} {D58}},\ \bibinfo {pages} {092003}
  (\bibinfo {year} {1998})},\ \Eprint {http://arxiv.org/abs/hep-ex/9712029}
  {arXiv:hep-ex/9712029 [hep-ex]} \BibitemShut {NoStop}%
\bibitem [{\citenamefont {Mangano}\ and\ \citenamefont
  {Melia}(2015)}]{Mangano:2014xta}%
  \BibitemOpen
  \bibfield  {author} {\bibinfo {author} {\bibfnamefont {M.}~\bibnamefont
  {Mangano}}\ and\ \bibinfo {author} {\bibfnamefont {T.}~\bibnamefont
  {Melia}},\ }\href {\doibase 10.1140/epjc/s10052-015-3482-x} {\bibfield
  {journal} {\bibinfo  {journal} {Eur. Phys. J.}\ }\textbf {\bibinfo {volume}
  {C75}},\ \bibinfo {pages} {258} (\bibinfo {year} {2015})},\ \Eprint
  {http://arxiv.org/abs/1410.7475} {arXiv:1410.7475 [hep-ph]} \BibitemShut
  {NoStop}%
\bibitem [{\citenamefont {Grossman}\ \emph {et~al.}(2015)\citenamefont
  {Grossman}, \citenamefont {K{\"o}nig},\ and\ \citenamefont
  {Neubert}}]{Grossmann:2015lea}%
  \BibitemOpen
  \bibfield  {author} {\bibinfo {author} {\bibfnamefont {Y.}~\bibnamefont
  {Grossman}}, \bibinfo {author} {\bibfnamefont {M.}~\bibnamefont {K{\"o}nig}},
  \ and\ \bibinfo {author} {\bibfnamefont {M.}~\bibnamefont {Neubert}},\ }\href
  {\doibase 10.1007/JHEP04(2015)101} {\bibfield  {journal} {\bibinfo  {journal}
  {JHEP}\ }\textbf {\bibinfo {volume} {04}},\ \bibinfo {pages} {101} (\bibinfo
  {year} {2015})},\ \Eprint {http://arxiv.org/abs/1501.06569} {arXiv:1501.06569
  [hep-ph]} \BibitemShut {NoStop}%
\bibitem [{\citenamefont {Antusch}\ and\ \citenamefont
  {Fischer}(2015)}]{Antusch:2015mia}%
  \BibitemOpen
  \bibfield  {author} {\bibinfo {author} {\bibfnamefont {S.}~\bibnamefont
  {Antusch}}\ and\ \bibinfo {author} {\bibfnamefont {O.}~\bibnamefont
  {Fischer}},\ }\href {\doibase 10.1007/JHEP05(2015)053} {\bibfield  {journal}
  {\bibinfo  {journal} {JHEP}\ }\textbf {\bibinfo {volume} {05}},\ \bibinfo
  {pages} {053} (\bibinfo {year} {2015})},\ \Eprint
  {http://arxiv.org/abs/1502.05915} {arXiv:1502.05915 [hep-ph]} \BibitemShut
  {NoStop}%
\end{thebibliography}%

\end{document}